\documentclass[aps,prc,superscriptaddress,groupedaddress,showpacs]
{revtex4}
\usepackage{graphicx}
\begin{document}
\title{Two-proton overlap functions in the Jastrow correlation method
and cross section of the $^{16}$O$(e,e^{\prime}pp)^{14}$C$_{\rm g.s.}$
reaction}

\author{D.~N.~Kadrev}
\affiliation{Institute for Nuclear Research and Nuclear Energy,
Sofia 1784, Bulgaria} \affiliation{Istituto Nazionale di Fisica
Nucleare, Sezione di Pavia, Pavia, Italy}

\author{M.~V.~Ivanov}
\affiliation{Institute for Nuclear Research and Nuclear Energy,
Sofia 1784, Bulgaria}

\author{A.~N.~Antonov}
\affiliation{Institute for Nuclear Research and Nuclear Energy,
Sofia 1784, Bulgaria}

\author{C.~Giusti}
\affiliation{Istituto Nazionale di Fisica Nucleare, Sezione di
Pavia, Pavia, Italy} \affiliation{Dipartimento di Fisica Nucleare
e Teorica, Universit\`a di Pavia, Pavia, Italy}

\author{F.~D.~Pacati}
\affiliation{Istituto Nazionale di Fisica Nucleare, Sezione di
Pavia, Pavia, Italy} \affiliation{Dipartimento di Fisica Nucleare
e Teorica, Universit\`a di Pavia, Pavia, Italy}

\begin{abstract}
Using the relationship between the two-particle overlap functions
(TOF's) and the two-body density matrix (TDM), the TOF's for the 
$^{16}$O$(e,e^{\prime}pp)^{14}$C$_{\rm g.s.}$ reaction are calculated
on the basis of a TDM obtained within the Jastrow correlation method.
The main contributions of the removal of $^1S_0$ and $^3P_1$ $pp$ pairs
from $^{16}$O are considered in the calculation of the cross section of
the $^{16}$O$(e,e^{\prime}pp)^{14}$C$_{\rm g.s.}$ reaction using the
Jastrow TOF's which include short-range correlations (SRC). The
results are compared with the cross sections calculated with different
theoretical treatments of the TOF's. 
\end{abstract}
\maketitle

\section{Introduction}

Since a long time electromagnetically induced two-nucleon knockout has
been devised as the most direct tool to study the properties of
nucleon pairs within nuclei at short distance and thus the dynamical
SRC in a nucleus \cite{Gottfried,Oxford}. 

Two nucleons can be naturally ejected by two-body currents, which
effectively take into account the influence of subnuclear degrees of
freedom like mesons and isobars. Direct insight into SRC can be
obtained from the process where the real or virtual photon hits,
through a one-body current, either nucleon of a correlated pair and
both nucleons are then ejected from the nucleus. The role and
relevance of these two competing processes can be different in
different reactions and kinematics. It is thus possible to envisage
situations where either process is dominant and various specific
effects can be disentangled and separately investigated.

Various theoretical models for cross section calculations have been
developed in recent years in order to explore the effects of
ground-state $NN$ correlations on $(e,e^{\prime}NN)$
\cite{GP91,RVH+95,RVH+97,GP97,GPA+98,GPM+99}) and $(\gamma,NN)$
\cite{GPR92,GP93,RVM+94,RMV+94,Ryc96,ic,GP98} knockout reactions. It
appears from these studies that the most promising tool for 
investigating SRC in nuclei is represented by the $(e,e^{\prime}pp)$
reaction, where the effect of the two-body currents is less dominant
as compared to the $(e,e^{\prime}pn)$ and $(\gamma,NN)$ processes.
Measurements of the exclusive $^{16}$O$(e,e'pp)^{14}$C reaction
performed at NIKHEF in Amsterdam \cite{Ond+97,Ond+98,Ronald} and MAMI
in Mainz~\cite{Ros99,Ros00} have confirmed, in comparison with the
theoretical results, the validity of the direct knockout mechanism for
transitions to low-lying states of the residual nucleus and have given
clear evidence of SRC for the transition to the ground state of
$^{14}$C. This result opens up good perspectives that further
theoretical and experimental efforts on two-nucleon knockout will be
able to determine SRC. 

One of the main ingredients in the transition matrix elements of
exclusive two-nucleon knockout reactions is the two-nucleon overlap
function. The TOF contains information on nuclear structure and
correlations and allows one to write the cross section in terms of the
two-hole spectral function \cite{Oxford}. The TOF's and their
properties are widely reviewed, e.g., in \cite{BGP+85}. In an
inclusive reaction, integrating the spectral function over the whole
energy spectrum produces the TDM.

In \cite{GP97} the TOF's for the $^{16}$O$(e,e'pp)^{14}$C reaction are
given by the product of a coupled and fully antisymmetrized pair
function of the shell model and a Jastrow-type correlation function
which incorporates SRC. Only the central term of the correlation
function is retained in the calculation. 

A more sophisticated treatment is used in \cite{GPA+98}, where the
TOF's are obtained from an explicit calculation of the two-proton
spectral function of $^{16}$O \cite{GAD+96}, which includes, with some
approximations but consistently, both SRC and long-range correlations
(LRC). The numerical predictions of this model are in reasonable and
in some cases in good agreement with data
\cite{Ond+98,Ronald,Ros99,Ros00}. 

Although satisfactory, these first comparisons with data have also
raised problems that require further theoretical investigation and a
more refined treatment of nuclear structure aspects in the calculation
of the spectral function. The study of the two-hole spectral functions
including different types of correlations, however, requires
substantial efforts in computational many-body physics and represents
a very difficult task.

A different method to calculate the TOF's has been suggested in
\cite{ADS+99} using the established general relationships connecting
TOF's with the ground state TDM. The procedure is based on the
asymptotic properties of the TOF's in coordinate space, when the
distance between two of the particles and the center of mass of the
remaining core becomes very large. This procedure can be considered as
an extension of the method suggested in \cite{NWH93}, where the
relationship between the one-body density matrix and the one-nucleon
overlap function is established. The latter has been applied
\cite{SAD96,NDW97,NDD+96,DGA+97,GPD97,GPA00,IGA01,IGA02} to calculate
the one-nucleon overlap functions, spectroscopic factors and to make
consistent calculations of the cross sections of one-nucleon removal
reactions $(p,d)$, $(e,e^{\prime}p)$, and $(\gamma,p)$
\cite{SAD96,DGA+97,GPD97,GPA00,IGA01,IGA02} on $^{16}$O
\cite{GPD97,GPA00,IGA02} and $^{40}$Ca \cite{IGA02,IGA01}. Various
correlation methods, such as the Jastrow method, the Green function
method, the correlated basis function method and the generator
coordinate method have been used to obtain the one-nucleon overlap
functions which are necessary for cross section calculations.

The first aim of the present paper is to apply the procedure suggested
in \cite{ADS+99} to calculate TOF's for $^{16}$O using the TDM
calculated in \cite{DKA+00} with the Jastrow correlation method (JCM),
which incorporates the nucleon-nucleon SRC. As a second aim, the
resulting two-proton overlap functions are used to calculate the cross
section of the $^{16}$O$(e,e^{\prime}pp)$ reaction for the transition
to the ground state of $^{14}$C. The cross sections are calculated on
the basis of the theoretical approach developed in
\cite{GP91,GP97,GPA+98}. The choice of the Jastrow TDM is determined
by the convenience of its analytical form obtained in \cite{DKA+00},
which makes practically possible the calculation of the TOF's. Of
course, the reliability of the TOF's obtained in our method depends
strongly on the availability of realistic TDM's. So, the usage in our
work of the Jastrow TDM, though incorporating only SRC (and using
harmonic-oscillator single-particle wave functions in the Slater
determinant), must be considered as a first attempt to use an approach
which fulfils the general necessity the TOF's to be extracted from the
TDM and to apply them to cross section calculations of two-nucleon
knockout reactions.

The method to calculate the TOF's on the basis of the TDM is briefly
outlined in Section~II. The results of the calculations of the TOF's
and the cross section of the $^{16}$O$(e,e^{\prime}pp)^{14}$C$_{\rm
g.s.}$ reaction are presented and discussed in Section~III. Some
conclusions are drawn in Section~IV.

\section{two-body density matrix and overlap functions}

In this Section we present shortly the definitions and some properties
of the TDM and related quantities in both natural orbital (geminal) and
overlap function representations. The method to extract the TOF's from
the TDM \cite{ADS+99} used in this work is also given.

The TDM is defined in coordinate space as:
\begin{equation}
\rho ^{(2)}(x_{1},x_{2};x_{1}^{\prime},x_{2}^{\prime})=\langle
{\Psi } ^{(A)}|a^{\dagger}(x_{1}) a^{\dagger}(x_{2}) a(x_{2}^{\prime})
a(x_{1}^{\prime})|{ \Psi }^{(A)}\rangle , \label{tdm}
\end{equation}
where $|{\Psi }^{(A)}\rangle $ is the antisymmetric $A$-fermion
ground state wave function normalized to unity and $a^{\dagger}(x)$,
$a(x)$ are creation and annihilation operators at position $x$.
The coordinate $x$ includes the spatial coordinate ${\bf r}$ and spin
and isospin variables. The TDM $\rho ^{(2)}$
is trace-normalized to the number of pairs of particles:
\begin{equation}
{\it Tr} \rho ^{(2)}=\frac{1}{2} \int \rho ^{(2)}(x_{1},x_{2})
dx_{1} dx_{2}\;=\;\frac{A(A-1)}{2} . \label{norm2}
\end{equation}

Since $\rho ^{(2)}$ is a Hermitian matrix its eigenstates
$\psi_{\alpha }^{(2)}$ form a complete orthonormal set in terms of
which $\rho ^{(2)}$ can be decomposed as
\begin{equation}
\rho^{(2)} (x_1,x_2;x_1^{\prime},x_2^{\prime})=\sum\limits_{\alpha
} \lambda _{\alpha}^{(2)}\psi _{\alpha}^{(2)*}(x_1,x_2)\psi
^{(2)}_{\alpha}(x_1^{\prime},x_2^{\prime})\;. \; \label{ro_gem}
\end{equation}
The eigenfunctions $\psi _{\alpha }^{(2)}(x_{1},x_{2})$ are called
natural geminals and the associated real eigenvalues $\lambda
_{\alpha }^{(2)}$ -- natural geminal occupation numbers
\cite{Pea75}. As a consequence of the antisymmetry of the nuclear
ground state, the eigenvalues $\lambda _{\alpha }^{(2)}$ obey the
inequalities:
\begin{equation}
\begin{array}{lll}
0\;\leq \lambda _{\alpha }^{(2)} \leq (A-1)/2 & \ \ \ &{\rm for}\;\;
A\;\;
{\rm odd,}\\
0\;\leq \lambda _{\alpha }^{(2)} \leq A/2 & &{\rm for}\;\; A \;\; {\rm
even. } \label{ineq2}
\end{array}
\end{equation}

The upper bound in Eq. (\ref{ineq2}) is actually reached only for
systems which are maximally correlated, as, e.g., the occupation number
of zero-coupled pairs in the seniority formalism in the limit of large
shell degeneracy.

Of direct physical interest is the decomposition of the TDM in terms of
the overlap functions between the $A$-particle ground state and the
eigenstates of the $(A-2)$-particle systems, since TOF's can be probed
in exclusive knockout reactions.

The TOF's are defined as the overlap between the ground state of the
target nucleus ${\Psi }^{(A)}$ and a specific state ${\Psi}
_{\alpha}^{(C)}$ of the residual nucleus ($C=A-2$) \cite{BGP+85}:
\begin{equation}
\Phi _{\alpha }(x_{1},x_{2})=\langle {\Psi }_{\alpha
}^{(C)}|a(x_{1}) a(x_{2})|{\Psi }^{(A)}\rangle . \label{of2}
\end{equation}

Inserting a complete set of $(A-2)$ eigenstates $|\alpha (A-2)\rangle $
into Eq.~(\ref{tdm}) one gets
\begin{equation}
\rho ^{(2)}(x_{1},x_{2};x^{\prime}_{1},x^{\prime}_{2}) =\sum
\limits_{\alpha } \Phi _{\alpha }^{*}(x_{1},x_{2}) \Phi _{\alpha
}(x^{\prime}_{1},x^{\prime}_{2}) . \label{ro_of}
\end{equation}

The norm of the two-body overlap functions defines the spectroscopic
factors
\begin{equation}
S_{\alpha }^{(2)}=\langle \Phi _{\alpha }|\Phi _{\alpha }\rangle .
\label{sf2}
\end{equation}

As in the case of the single-particle spectroscopic factors, where the
latter cannot exceed the maximal natural occupation number
\cite{NWH93}, one can find that $S_{\alpha }^{(2)} \leq \lambda
_{max}^{(2)}$.

A procedure for obtaining the TOF's on the basis of the TDM has been
suggested in \cite{ADS+99}. It is due to the particular asymptotic
properties of the TOF's and is similar to the one suggested in
\cite{NWH93} for deriving the one-body overlap functions from the
one-body density matrix. 

In the case when two like nucleons (neutrons or protons) unbound to the
rest of the system are simultaneously transferred, the following
hyperspherical type of asymptotics is valid for the two-body overlap
functions \cite{BGP+85,Mer74,Ban80}
\begin{equation}
\Phi (r,R) \longrightarrow N \exp \left\{
-\sqrt{\frac{4m|E|}{\hbar ^{2}} \left(
R^{2}+\frac{1}{4}r^{2}\right) }\right\} \left( R^{2}+\frac{1}{4}
r^{2}\right) ^{-5/2}\; , \label{of2as}
\end{equation}
where $r$ and $R$ are the magnitudes of the relative and center-of-mass
(CM) coordinates, ${\bf r}={\bf r}_{1}-{\bf r}_{2}$ and  ${\bf
R}=({\bf r}_{1}+{\bf r} _{2})/2$, respectively, $m$ is the nucleon
mass and $E=E^{(A)}-E^{(C)}$ is the two-nucleon separation energy.

For a target nucleus with $J^{\pi }_{tar.}=0^{+}$ the TOF in
Eq.~(\ref{of2}) can be written in the form
\begin{equation}
\Phi _{\nu JM}(x_{1},x_{2})=\sum_{LS}\left\{ \Phi _{\nu JLS}({\bf
r}_{1}, {\bf r}_{2}){\otimes }\chi _{S}(\sigma _{1},\sigma
_{2})\right\} _{JM} , \label{S_dec}
\end{equation}
where $\nu $ is the number of the state of the residual nucleus
with a given total momentum $J$,
\begin{equation}
\chi_{SM_{S}}(\sigma _{1},\sigma _{2}) = \left\{ \chi
_{\frac{1}{2} }(\sigma _{1})\otimes \chi _{\frac{1}{2}}(\sigma
_{2})\right\} _{SM_{S}} =\sum\limits_{m_{s_{1}}m_{s_{2}}}\left(
\frac{1}{2} m_{s_{1}} \frac{1}{2} m_{s_{2}}|SM_{S}\right) \chi
_{\frac{1}{2}m_{s_{1}}}(\sigma _{1}) \chi _{
\frac{1}{2}m_{s_{2}}}(\sigma _{2}), \label{S_fun}
\end{equation}
and $\Phi _{\nu JLSM_{L}}({\bf r}_{1},{\bf r}_{2})$ is the spatially
dependent part of the overlap function. Performing a decomposition into
angular momenta ${\mathbf l}={\bf l}_{r} $ and ${\bf L}_{R}$ (${\bf
L}={\bf l}+{\bf L}_{R}$) corresponding to the relative and CM
coordinates one obtains:
\begin{equation}
\Phi _{\nu JSLM_{L}}({\bf r},{\bf R})=\sum_{lL_{R}}\Phi _{\nu
JSLlL_{R}}(r,R)\left\{ Y_{L_{R}}(\widehat{R})\otimes Y_{l}(
\widehat{r})\right\} _{LM_{L}}\;. \label{L_dec}
\end{equation}

Then the TDM can be written as
\begin{equation}
\rho ^{(2)}(x_{1},x_{2};x_{1}^{\prime},x_{2}^{\prime}) = \sum_{JM}
\mathop{\sum_{LS}}_{L^{\prime}S^{\prime
}}\mathop{\sum_{lL_{R}}}_{l^{\prime}L_{R}^{\prime
}}\mathop{\rho^{(2)}_{JSLlL_{R}}}_{\;\;\;\;\;\;S^{\prime}L^{\prime
}l^{\prime}L_{R}^{\prime}}(r,R;r^{\prime},R^{\prime}) \;
A_{SLlL_{R}}^{JM *}(\sigma _{1},\sigma _{2};\widehat{r},
\widehat{R}) \; A_{S^{\prime}L^{\prime}l^{\prime}L_{R}^{\prime
}}^{JM}(\sigma _{1}^{\prime},\sigma
_{2}^{\prime};\widehat{r}^{\prime},\widehat{R}^{\prime}) ,
\label{tddec}
\end{equation}
where the radial part of the density matrix is
\begin{equation}
\mathop{\rho^{(2)}_{ JSLlL_{R}}}_{\;\;\;\;\;\;S^{\prime}L^{\prime
}l^{\prime}L_{R}^{\prime}}(r,R;r^{\prime},R^{\prime})=\sum_{\nu
}\Phi _{\nu JSLlL_{R}}^{*}(r,R) \Phi _{\nu
JS^{\prime}L^{\prime}l^{\prime
}L_{R}^{\prime}}(r^{\prime},R^{\prime}) \label{ror}
\end{equation}
and the spin-angular function is
\begin{equation}
A_{SLlL_{R}}^{JM}(\sigma _{1},\sigma _{2};\widehat{r},\widehat{R}
)=\left\{ \left\{ Y_{L_{R}}(\widehat{R})\otimes Y_{l}(\widehat{r}
)\right\} _{LM_{L}}\otimes \chi _{SM_{S}}(\sigma _{1},\sigma
_{2})\right\} _{JM}. \label{tdso}
\end{equation}

We will consider the diagonal part of the radial TDM in Eq.~(\ref{ror}): 
\begin{equation}
\rho _{JSL{l}L_{R}}^{(2)}(r,R;r^{\prime},R^{\prime})= \sum_{\nu
}\Phi _{\nu JSLlL_{R}}^{*}(r,R) \Phi _{\nu JSLlL_{R}}(r^{\prime
},R^{\prime}). \label{rord}
\end{equation}

For large $r^{\prime}=a$ and $R^{\prime}=b$ a single term with $\nu_0$,
corresponding to the smallest two-nucleon separation energy, will
dominate the sum in the right-hand side of Eq.~(\ref{rord}). Then,
according to Eq.~(\ref{of2as}), the radial part of the TOF $\Phi
_{\nu_0 JSLlL_{R}}(r,R)$ can be expressed in terms of the TDM as
\begin{equation}
\Phi _{\nu_0 JSLlL_{R}}(r,R) = \frac{\rho
_{JSLlL_{R}}^{(2)}(r,R;a,b)
}{\Phi _{\nu_0 JSLlL_{R}}(a,b)} %\\
= \frac{\rho _{JSLlL_{R}}^{(2)}(r,R;a,b)}{N \exp \left\{
-k\sqrt{\left( b^{2}+\frac{1}{4}a^{2}\right) }\right\} \left(
b^{2}+\frac{1}{4} a^{2}\right) ^{-5/2}} ,
\label{p2of}
\end{equation}
where $k=(4m|E|/\hbar ^{2} )^{1/2}$ is constrained by the experimental
values of the two-nucleon separation energy $E$.

The relationship obtained in Eq.~(\ref{p2of}) makes it possible to
extract TOF's with quantum numbers $JSLlL_{R}$ from a given TDM. The
coefficient $N$ and the constant $k$ can be determined from the
asymptotics of ${\rho^{(2)}_{JSLlL_{R}}}(r,R;r,R)$.

\section{Results}

\subsection{The two-proton overlap functions}

The procedure described in Section II has been applied to calculate
the two-proton overlap functions in the $^{16}$O nucleus for the
transition to the $0^+$ ground state of $^{14}$C. The TDM obtained in
\cite{DKA+00} in the framework of the low-order approximation (LOA) of
the Jastrow correlation method has been used. In \cite{DKA+00} the
latter incorporates the nucleon-nucleon SRC in terms of the
wave-function ansatz \cite{Jas55}:
\begin{equation}
\Psi ^{(A)}({\bf r}_{1},{\bf r}_{2},\ldots ,{\bf
r}_{A})=(C_{A})^{-1/2} \prod_{1\leq i<j\leq A}f(\mid {\bf
r}_{i}-{\bf r}_{j}\mid ) \Phi _{SD}^{A}( {\bf r}_{1},{\bf
r}_{2},\ldots ,{\bf r}_{A}), \label{jas}
\end{equation}
where $C_{A}$ is a normalization constant and $\Phi _{SD}^{A}$ is a
single Slater determinant wave function built from harmonic oscillator
(HO) single-particle wave functions which depend on the oscillator
parameter $\alpha_{osc.}$, having the same value for both protons and
neutrons. Only central correlations are included in the correlation
factor $f(r)$, which is state-independent and has a simple Gaussian
form
\begin{equation}
f(r) = 1-c\ \exp (-\beta ^{2}r^{2}), 
\label{fcorr} \end{equation}
where the correlation parameter $\beta $ determines the healing
distance and the parameter $c$ accounts for the strength of the SRC. 
The LOA keeps all terms up to the second order in $h=f-1$ and the first
order in $g=f^{2}-1$ in such a way that the normalization of the
density matrices is ensured order by order \cite{GGR71}.

The values of parameters $\alpha_{osc.}$ and $\beta $ have been
obtained \cite{SAD93} phenomenologically by fitting the experimental
elastic formfactor data for $^4$He, $^{16}$O and $^{40}$Ca nuclei. The
value of the parameter $c$ has been determined \cite{DKA+00} under the
additional condition the relative pair density distribution
\begin{equation}
\rho ^{(2)}({\bf r)=}\int \rho ^{(2)}({\bf r},{\bf R};{\bf r},{\bf
R})d{\bf R} \label{denrelr}
\end{equation}
to reproduce at $r=0$ the associated value obtained within the
Variational Monte-Carlo approach \cite{PWP92}. Thus, in the present
calculations the following values of the parameters are used for
$^{16}$O: $\alpha_{osc.} =0.61\;{\rm fm^{-1}}$, $\beta =1.30\;{\rm
fm^{-1}}$, $c=0.77$.

In order to obtain the radial part of the TDM, $\rho^{(2)} (r,R;
r^{\prime}, R^{\prime})$ of Eq.~(\ref{ror}), we use the analytical
expression for the TDM obtained in \cite{DKA+00} substituting the 
coordinates of the two particles, ${\bf r}_{1}$ and ${\bf r}_{2}$, by
the CM ${\bf R}$ and relative ${\bf r}$ coordinates. Then, the TDM is 
multiplied by $A_{S^{\prime\prime} L^{\prime\prime} l^{\prime\prime}
L_{R} ^{\prime\prime}}^{JM}(\sigma _{1},\sigma
_{2};\widehat{r},\widehat{R} ) A_{S^{\prime\prime\prime}
L^{\prime\prime\prime} l^{\prime\prime\prime}
L_{R}^{\prime\prime\prime}} ^{JM*} (\sigma_{1}^{\prime},
\sigma_{2}^{\prime};\widehat{r}^{\prime},\widehat{R}^{\prime} )$ and
the integration over the angles and summation over the spin variables
lead to the radial part $\displaystyle
\mathop{\rho^{(2)}_{JS^{\prime\prime} L^{\prime\prime} l^{\prime\prime}
L^{\prime\prime}_{R}}} _{\;\;\;\;\;\;S^{\prime\prime\prime}
L^{\prime\prime\prime} l^{\prime\prime\prime}
L_{R}^{\prime\prime\prime}} (r,R;r^{\prime},R^{\prime}).$

% The fitting procedure

In order to obtain the values of the parameters $k$ and $N$ in
Eq.~(\ref{p2of}) simultaneously, we look for such a radial contribution
$\rho_{\nu_0JSLlL_{R}}^{(2)}(r,R;r',R')= \Phi_{\nu_0 JSLlL_{R}}
^{*}(r,R) \Phi_{\nu_0 JSLlL_{R}}(r^{\prime },R^{\prime})$ whose
diagonal part $\rho_{\nu_0 JSLlL_{R}}^{(2)}(r,R)$ minimizes the trace
\begin{equation}
{\rm Tr}\left[ \left(\rho_{JSLlL_{R}}^{(2)} (r,R) - \rho_{\nu_0
JSLlL_{R}}^{(2)}(r,R) \right)^2 \right]= \min . \label{Tr}
\end{equation}

The correct determination of these parameters requires a proper
definition of the asymptotic region where the trace in Eq.~(\ref{Tr})
has to be minimized. If we denote the point in which $\rho
_{JSLlL_{R}}^{(2)}(r,R)$ has a maximum with $(r_{max},R_{max})$, the
starting point of the asymptotic region $(r_0,R_0)$ is obtained looking
for a point $r_0$, at $R=R_{max}$, for which $\rho _{JSLlL_{R}}^{(2)}
(r_0,R_{max}) \leq 10\%$ of $\rho _{JSLlL_{R}} ^{(2)}(r_{max},
R_{max})$. When $r_0$ has been determined, we look for a point $R_0$,
at $r=r_0$, for which  $\rho _{JSLlL_{R}}^{(2)}(r_0,R_0) \leq 10\%$ of
$\rho _{JSLlL_{R}}^{(2)}(r_0,R_{max})$. The length of the asymptotic
region over $r$ and $R$ is determined by the requirement to obtain the
separation energy which is maximally close to the experimental one. The
asymptotic point ($a,b$) is chosen to be that one which gives the
minimal least-squared deviation expressed by Eq.~(\ref{Tr}).

% Results OF

When all the parameters are determined, Eq.~(\ref{p2of}) can be used to
calculate the radial part $\Phi _{\nu_0 JSLlL_{R}}(r,R)$. Then,
including also the spin-angular part in Eqs.~(\ref{L_dec}) and
(\ref{S_dec}) we obtain the TOF's.

For a given set of quantum numbers $JSLlL_{R}$ the TOF is calculated 
by minimizing the trace of the corresponding part of the TDM. Thus,
for a particular final state $J^\pi$ of the residual nucleus, different
TOF's can be independently calculated using this procedure for each set
of quantum numbers, and each one of them is fully responsible for the
two-proton knockout process and the transition to the state
$J^{\pi}$. 

\begin{figure}
\begin{center}\begin{tabular}{ccc}
\includegraphics[width=7cm]{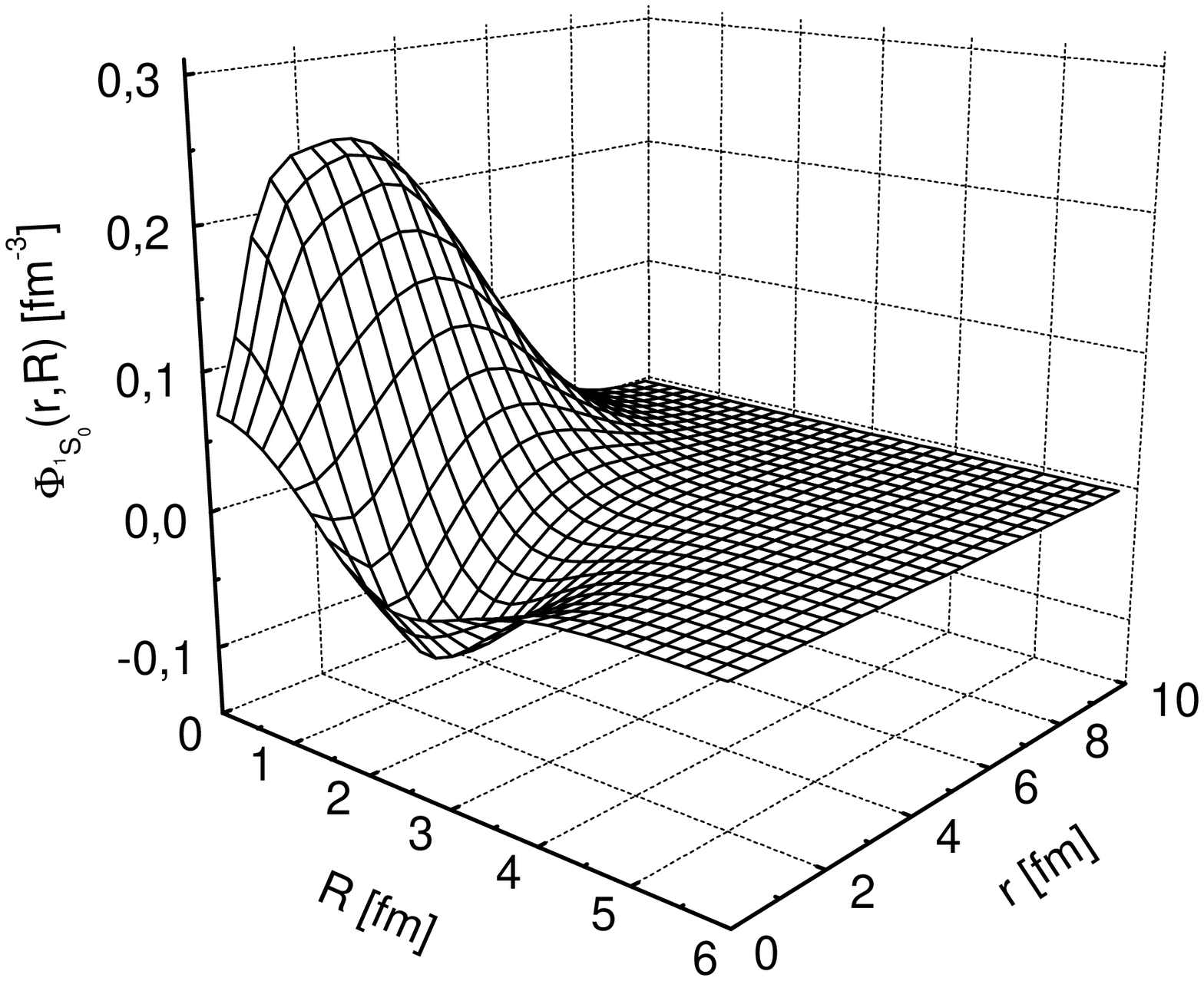} & &
\includegraphics[width=7cm]{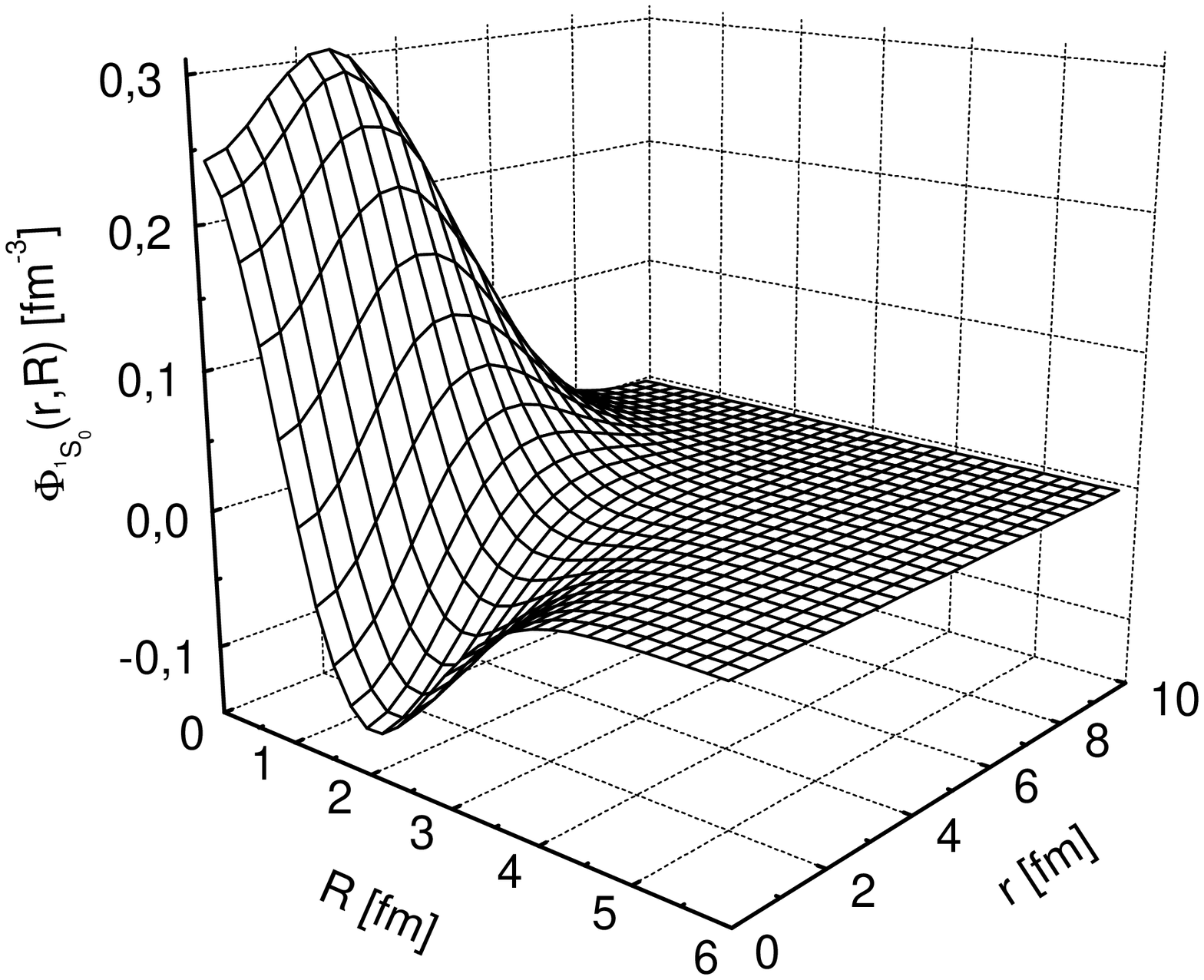}
\end{tabular}\end{center}
\caption{The $^1S_0$ two-proton overlap functions for the nucleus
$^{16}$O leading to the $0^+$ ground state of $^{14}$C extracted
from the JCM (left) and uncorrelated (right) two-body density
matrices. \label{f1S}}
\end{figure}

\begin{figure}
\begin{center}\begin{tabular}{ccc}
\includegraphics[width=7cm]{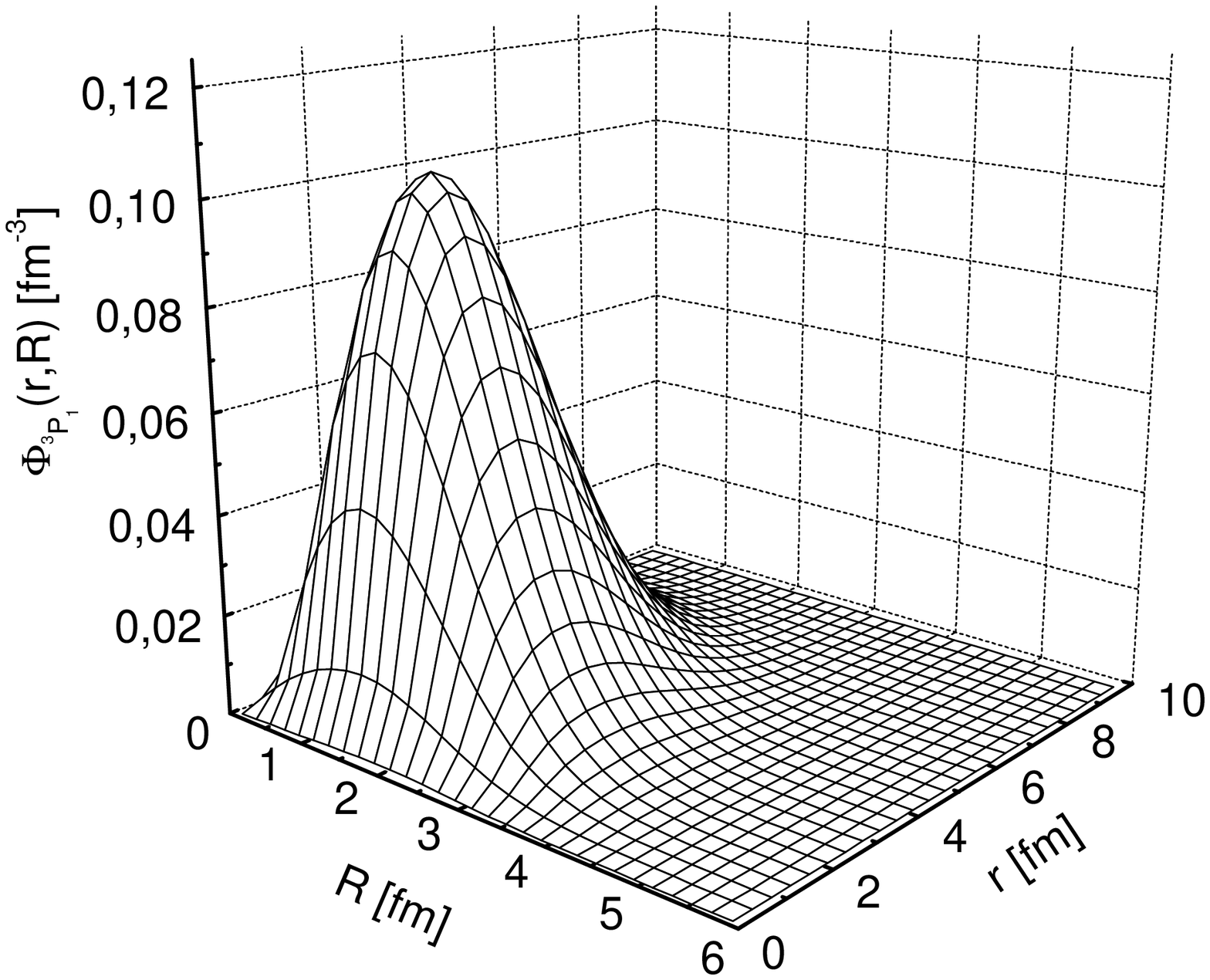} & &
\includegraphics[width=7cm]{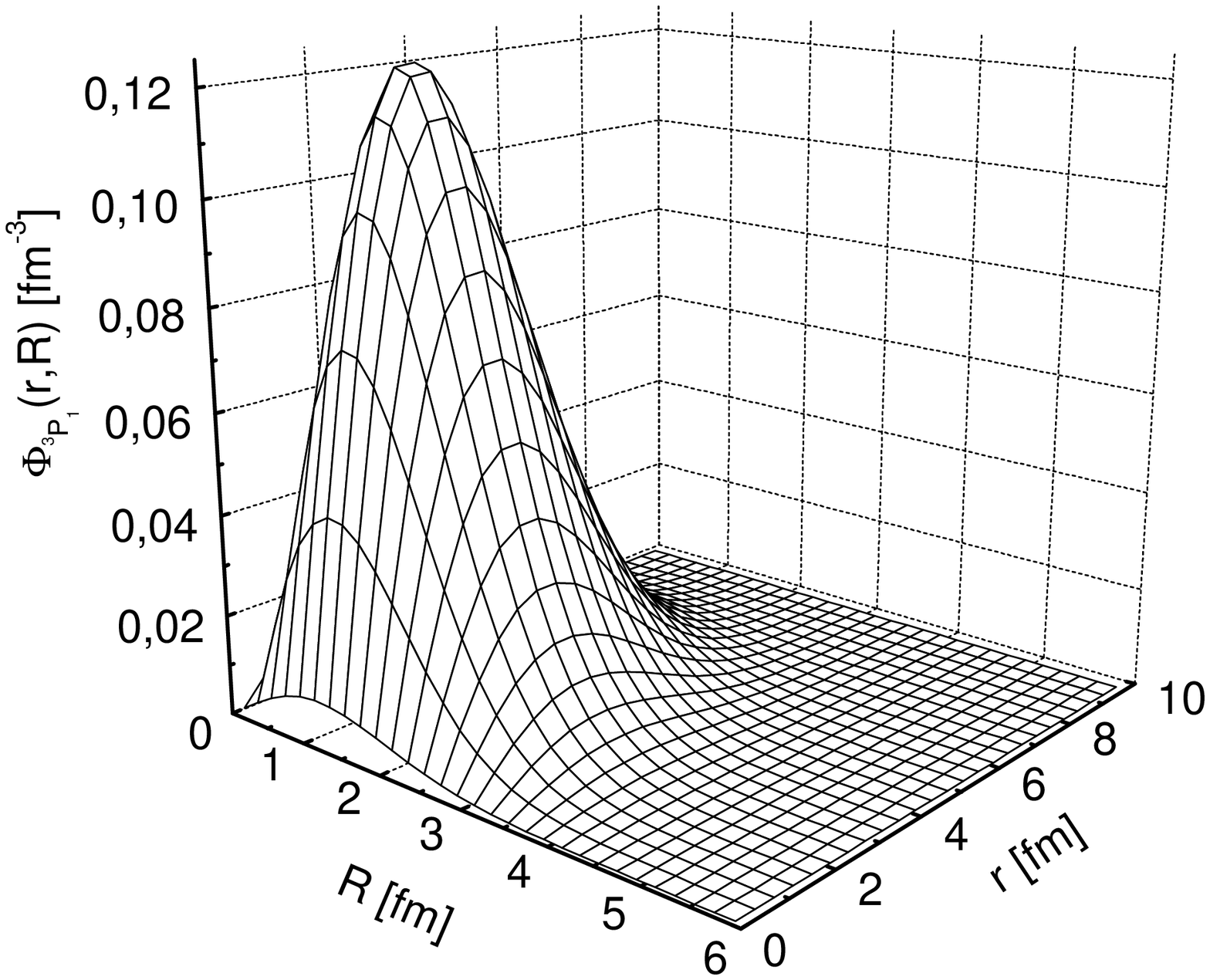}
\end{tabular}\end{center}
\caption{The $^3P_1$ two-proton overlap functions for the nucleus
$^{16}$O leading to the $0^+$ ground state of $^{14}$C extracted
from the JCM (left) and uncorrelated (right) two-body density
matrices. \label{f3P}}
\end{figure}

The TOF's obtained in the JCM for the $^1S_0$ and $^3P_1$ states are
presented in Figs.~\ref{f1S} and ~\ref{f3P}, respectively. They are
compared with the uncorrelated TOF's obtained applying the same
procedure to the uncorrelated TDM, i.e. with $c=0$ in the correlation 
factor of Eq.~(\ref{fcorr}). The notation for the partial waves in our
case is $^{2S+1}l_L$. It differs from the generally accepted one
$^{2S+1}l_j$ because we have a different coupling scheme of spin and
angular momenta.

SRC depend on both relative and CM coordinates and it can be seen from
the figures that they affect both size and shape of the TOF's. Their
role, however, is different in the two states and, as it was already
found in previous and different calculations (see, e.g., 
\cite{GAD+96,GPA+98}), are much more important when the two protons are
in the $^1S_0$ than in the $^3P_1$ state. 

The spectroscopic factors corresponding to the $^1S_0$ and $^3P_1$
overlap functions are 0.958 and 0.957, respectively. Also the $D$ 
wave can contribute for the transition to the $0^+$ ground state of
$^{14}$C, but the corresponding TOF is very small and is not 
considered in the present study.

% Total OF

As a next step, we derive the total TOF ${\Phi}_{\nu JM}({x},{X})$
in terms of a sum over different partial components, i.e.
\begin{equation}
{\Phi}_{\nu JM}({x},{X})=\sum_{LSlL_{R}}\Phi_{\nu JSLlL_{R}}(r,R)
A_{SLlL_{R}}^{JM }(\sigma _{1},\sigma _{2};\widehat{r},
\widehat{R}). \label{total}
\end{equation}

We integrate the squared modulus of the total TOF in Eq.~(\ref{total})
over the angles and sum over the spin variables. The result can be
written in the form (for the smallest value of $\nu=\nu_0$):
\begin{equation}
\overline{|{\Phi}_{JM}({x},{X})|^{2}}
{\equiv}|\widetilde{\Phi}_{JM}({r},{R})|^{2}
 =\sum_{LSlL_R}\rho^{(2)}_{JLSlL_R}(r,R), \label{21}
\end{equation}
where the bar denotes the integration over the angles and summation
over the spin variables, and $\widetilde{\Phi}_{JM}({r},{R})$ is the
radial part of the total TOF obtained after the integration and
summation. Using the asymptotics of $\widetilde{\Phi}_{JM}({r},{R})$
at $r \longrightarrow a$, $R\longrightarrow b$ one can write:
\begin{equation}
\widetilde{\Phi}_{JM}({r},{R})= \frac{\displaystyle \sum_{LSlL_{R}}\rho
_{JSLlL_{R}}^{(2)}(r,R;a,b)}{N \exp \left\{ -k\sqrt{\left(
b^{2}+\frac{1}{4}a^{2}\right) }\right\} \left( b^{2}+\frac{1}{4}
a^{2}\right) ^{-5/2}}\ \ .\label{22}
\end{equation}
The parameters $N$, $k$, $a$, $b$ in Eq.~(\ref{22}) can be
redetermined from the asymptotics of $\sum_{LSlL_R}{\rho^{(2)}
_{JLSlL_R}(r,R;r,R)}$ using the procedure already explained in the
first part of this Section. Then, each partial radial component
$\Phi_{JSLlL_{R}}(r,R)$ in Eq.~(\ref{total}) can be separately
calculated from Eq.~(\ref{p2of}) using for each one of them the same
coefficients $N$, $k$, $a$, $b$ which correspond to the asymptotics of
the total TOF. The asymptotic point ($a$,$b$) determines the
individual contribution of each partial overlap function to the total
TOF. This prescription allows us to combine, with some approximations,
the different radial components in Eq.~(\ref{total}). 

% Results total OF

The results for the $^1S_0$ and $^3P_1$ partial components have a
similar behaviour as in Figs.~\ref{f1S} and \ref{f3P}, the main
difference is that they are somewhat reduced in magnitude. The
reduction is determined by the contribution of each component to the
total TOF. The spectroscopic factor corresponding to the total TOF is
equal to unity in the uncorrelated case and 0.965 in the Jastrow case.

\subsection{The $^{16}$O$(e,e^{\prime}pp)^{14}$C$_{\rm g.s.}$ reaction}

% Cross sections - model

The TOF's obtained from the TDM within the Jastrow correlation method 
have been used to calculate the cross section of the 
$^{16}$O$(e,e^{\prime}pp)^{14}$C$_{\rm g.s.}$ knockout reaction. 

Calculations have been performed within the theoretical framework of 
\cite{GP91,GP97,GPA+98}. In this model the nuclear current operator is
the sum of a one-body and a two-body part. The one-body part contains
a Coulomb, a convective and a spin term. For $pp$ knockout the
two-body current contains only the contributions of non
charge-exchange processes with intermediate $\Delta$-isobar
configurations~\cite{GPA+98,ic}. In the final state the mutual
interaction between the two outgoing nucleons is neglected and the
scattering state is given by the product of two uncoupled 
single-particle distorted wave functions, eigenfunctions of a complex 
phenomenological optical potential which contains a central, a Coulomb
and a spin-orbit term \cite{Nad}. 

\begin{figure} \begin{center} \begin{tabular}{cc}
\includegraphics[width=7cm]{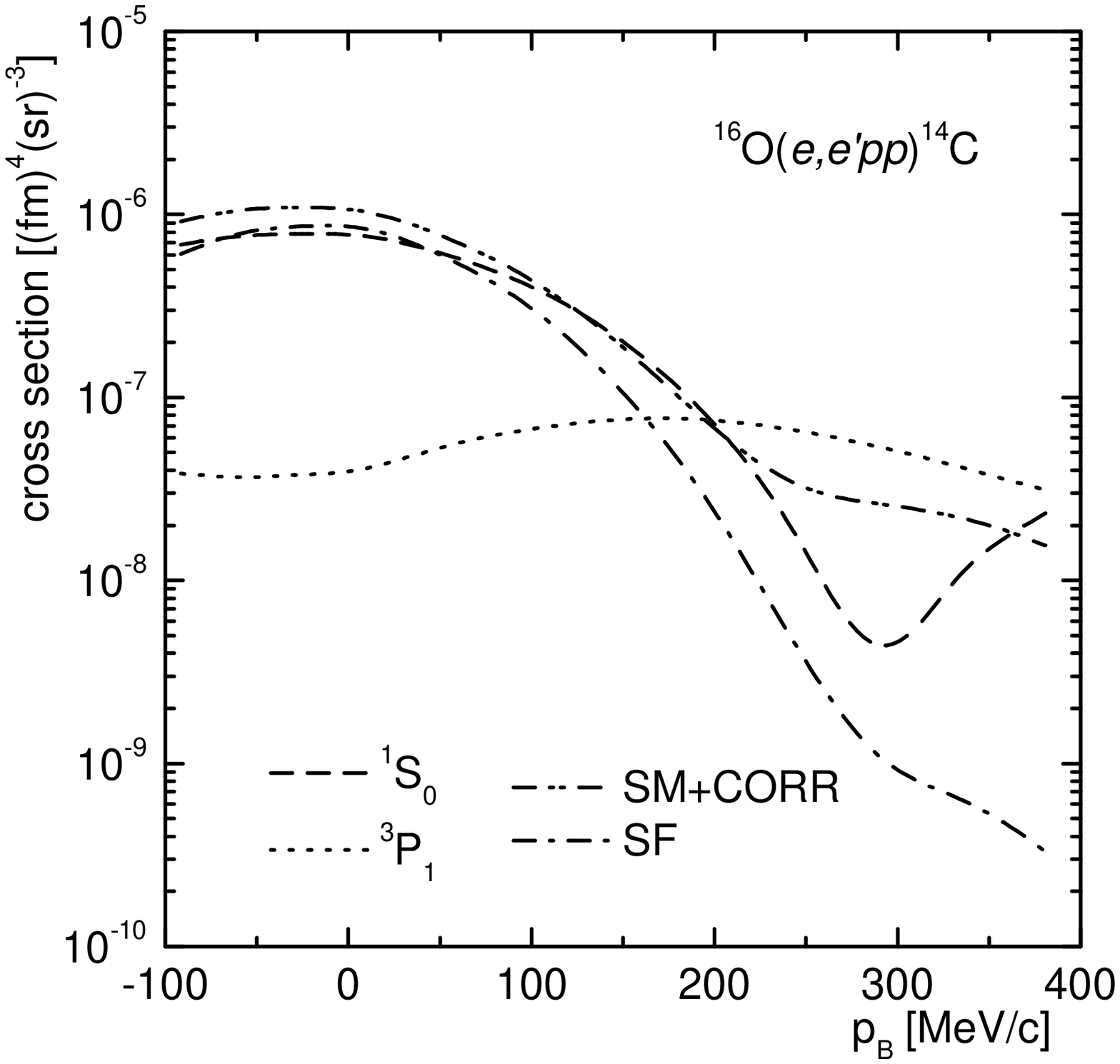} &
\includegraphics[width=7cm]{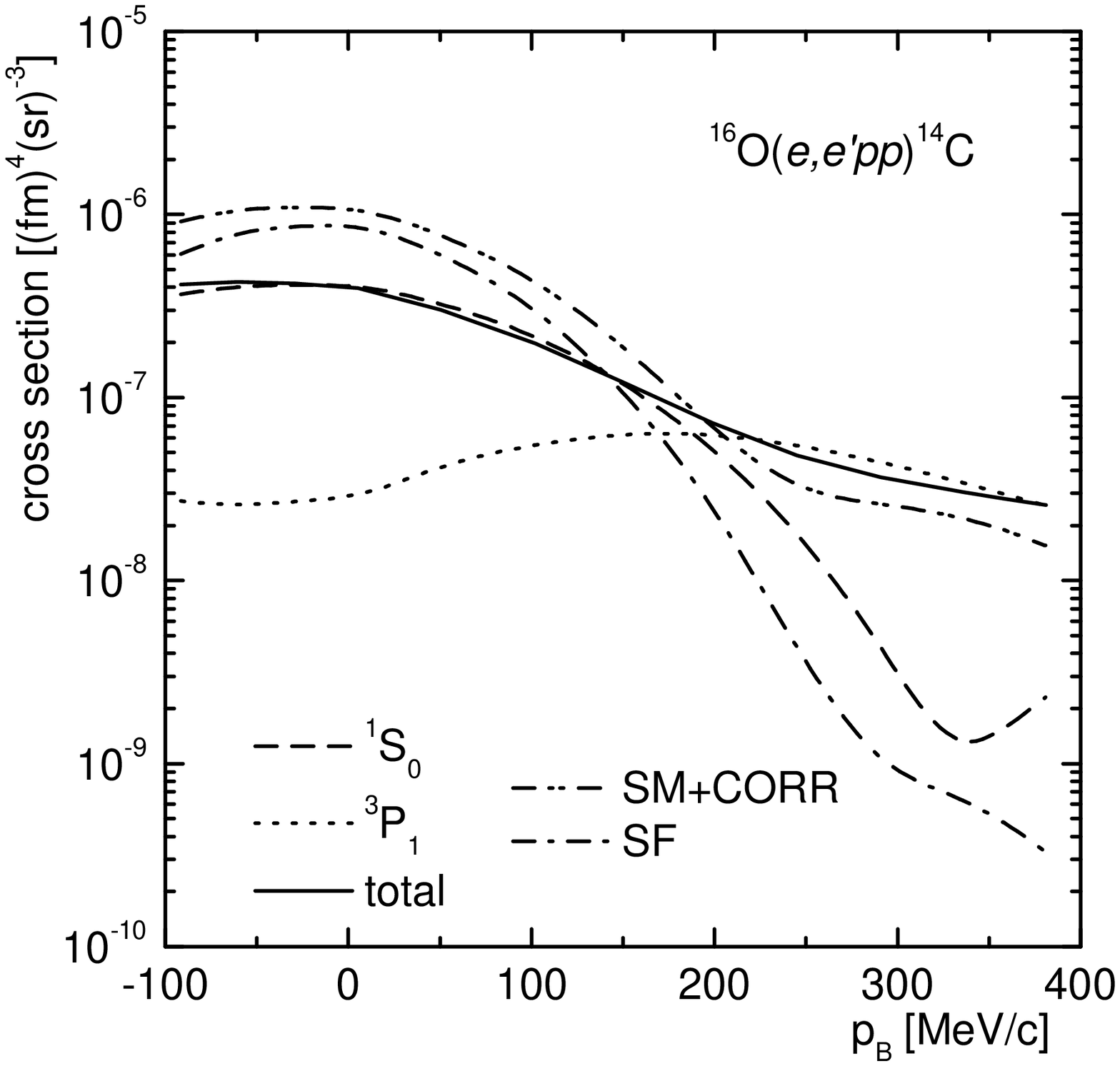} \end{tabular}\end{center}
\caption{The differential cross section of the $^{16}$O$(e,e^{\prime}p
p)^{14}$C$_{\rm g.s.}$ reaction as a function of the recoil momentum 
$p_{\rm B}$ in the superparallel kinematics with $E_0=855$ MeV, 
$\omega=215$ MeV and $q=316$ MeV/$c$. Positive (negative) values of
$p_{\rm B}$ refer to situations where ${\bf p}_{\rm B}$ is parallel
(anti-parallel) to ${\mbox{\boldmath $q$}}$. The curves are obtained
with different treatments of the TOF: $^1S_0$ (dashed line) and
$^3P_1$ (dotted line) as independent TOF's  in the JCM in the left
panel and as partial components in the right panel, the total TOF
(solid line), the TOF from the spectral function (SF)
\cite{GAD+96,GPA+98} (dot-dashed line), the product of a pair function
of the shell model and the correlation function of Eq.~(\ref{fcorr}) 
(SM+CORR) (dot-dot-dashed line). \label{cs_sup}}
\end{figure}

\begin{figure} \begin{center}\begin{tabular}{cccc}
\includegraphics[width=7cm]{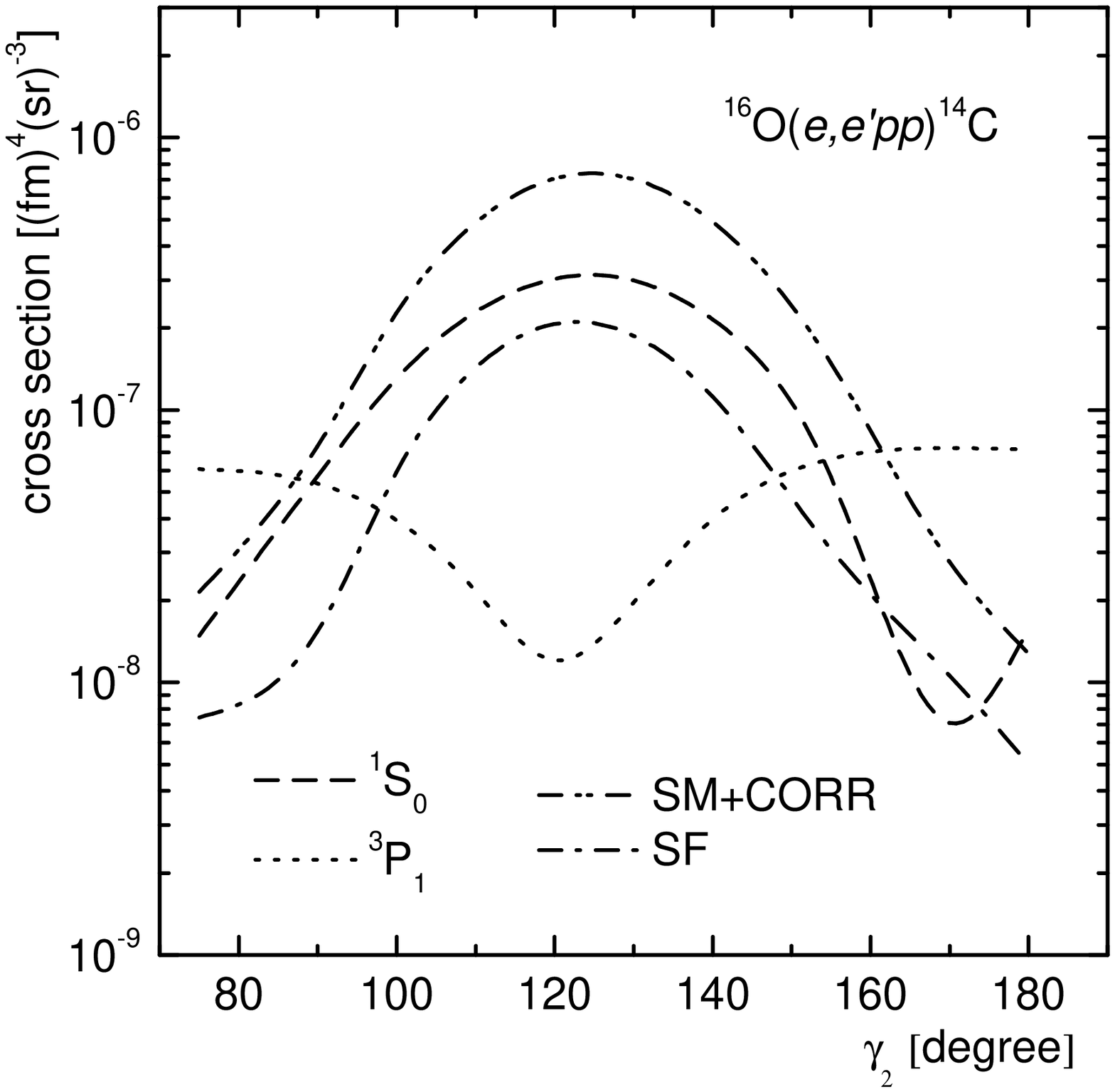} & & &
\includegraphics[width=7cm]{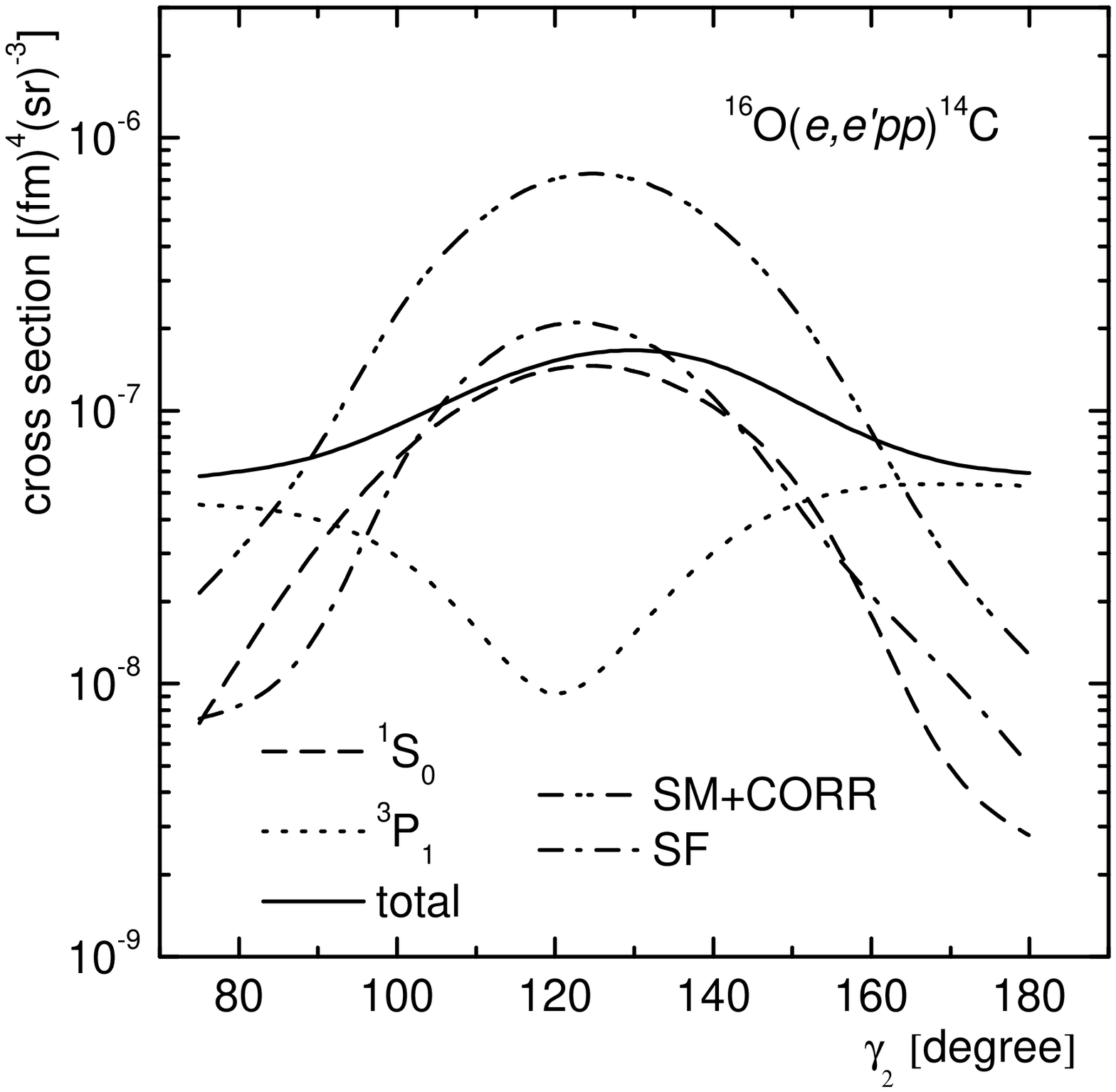} \end{tabular}\end{center}
\caption{The differential cross section of the $^{16}$O$(e,e^{\prime}
pp)^{14}$C$_{\rm g.s.}$ reaction as a function of the angle $\gamma_2$ 
in a NIKHEF kinematics with $E_0=584$ MeV, $\omega=212$ MeV, $q=300$
MeV/$c$, $T'_1=137$ MeV and $\gamma_1= -30^{\rm{o}}$, on the opposite
side of the outgoing electron with respect to the momentum transfer.
Line convention as in Fig.~\ref{cs_sup}. \label{cs_nik}} 
\end{figure}

Numerical results are shown in Figs.~\ref{cs_sup} and \ref{cs_nik} for
two kinematical settings considered in the experiments performed at
NIKHEF \cite{Ond+97,Ond+98,Ronald} and MAMI \cite{Ros99,Ros00}.  In
Fig.~\ref{cs_sup} the cross section is calculated in the
super-parallel kinematics of the MAMI experiment, where the two
nucleons are ejected parallel and anti-parallel to the momentum
transfer and, for a fixed value of the energy $\omega$ and momentum
transfer $q$, it is possible to explore, for different values of the
kinetic energies of the outgoing nucleons, all the possible values of
the recoil momentum $p_{\mathrm{B}}$ of the residual nucleus. In the 
calculations the incident electron energy is fixed at $E_0=855$ MeV, 
$\omega=215$ MeV and $q=316$ MeV/$c$. In Fig.~\ref{cs_nik} a specific 
kinematical setting included in the experiments carried out at NIKHEF
is considered, with $E_0=584$ MeV, $\omega = 212$ MeV and $q = 300$
MeV/$c$. The kinetic energy of the first outgoing proton $T'_1$ is 137
MeV and the angle $\gamma_1$, between the outgoing proton and ${\bf
q}$, is $30^{\mathrm{o}}$ on the opposite side of the outgoing
electron with respect to the momentum transfer. Changing the angle
$\gamma_2$ on the other side, different values of the recoil momentum
$p_{\mathrm{B}}$ are explored in the range between $-250$ and $300$
MeV/$c$, including the zero values at $\gamma_2 \simeq
120^{\mathrm{o}}$.

The cross sections calculated with the $^1S_0$ and $^3P_1$ TOF's as 
independent and fully responsible for the knockout process are
displayed in the left panels of Figs.~\ref{cs_sup} and \ref{cs_nik}.
In the right panels the cross sections obtained with the total TOF
from the Jastrow TDM are plotted and compared with the contributions
given by the $^1S_0$ and $^3P_1$ partial components. 

These results are compared in the figures with the cross sections
already shown in \cite{GPA+98}, where the TOF is taken from a
calculation of the two-proton spectral function (SF) \cite{GAD+96},
where a two-step procedure has been adopted to include both SRC and
LRC. LRC are calculated in a shell-model space large enough to
incorporate the corresponding collective features which influence the
pair removal amplitude. The single-particle propagators used for this
dressed random phase approximation (RPA) description of the
two-particle propagator also include the effect of both LRC and SRC. In
the second step that part of the pair removal amplitudes, which
describes the relative motion of the pair, is supplemented by defect
functions obtained from the same G-matrix which is also used as the
effective interaction in the RPA calculation. Different defect
functions are produced by different realistic $NN$ potentials. The
results shown in Figs.~\ref{cs_sup} and \ref{cs_nik} are obtained with
the Bonn-A potential. The explicit expression of the TOF's is given in
a form of the same kind as in Eq.~(\ref{total}), in terms of a
combination of CM and relative wave functions. The $^1S_0$ and $^3P_1$
relative waves give the main contribution for the transition to the
$0^+$ ground state, while only a negligible contribution is given by
the $D$ wave. The results of this model are able to give a proper 
description of available data \cite{Ond+98,Ronald,Ros99,Ros00}.

In the figures are also shown for a comparison the results obtained
with a simpler approach, where the two-nucleon wave function is given
by the product of the pair function of the shell model and of a
Jastrow type central and state independent correlation function. In
this approach (SM+CORR) the ground state of $^{14}$C is described as a
pure $(1p_{\frac{1}{2}})^{-2}$ hole in $^{16}$O. In order to allow a
more direct and clear comparison with the TOF's from the Jastrow TDM,
HO single-particle wave functions and the same correlation function as
in Eq.~(\ref{fcorr}), with the same parameters as in the calculation
of the TDM, have been adopted.

The shape of the calculated cross sections is determined by the value
of the CM orbital angular momentum $L_R$, that is $L_R=0$ for $^1S_0$
and $L_R=1$ for $^3P_1$. When the two components are combined in the
TOF the shape is driven by the component which gives the major
contribution, that is $L_R=0$ and $^1S_0$ at lower values and $L_R=1$
and $^3P_1$ at higher values of the recoil momentum.

The role of correlations and two-body currents is different in
different relative states. SRC are quite strong and even dominant for
the $^1S_0$ state and much weaker for the $^3P_1$ state. Moreover,
the role of the isobar current is strongly suppressed for $^1S_0$ $pp$
knockout, since there the generally dominant contribution of that
current, due to the magnetic dipole  $NN \leftrightarrow N\Delta$
transition, is suppressed \cite{GP98,delta}. Thus, the role of SRC is
emphasized in $^1S_0$ knockout, while the $\Delta$ current is
emphasized in $^3P_1$ knockout. These general features, which have
been found in all the previous studies of the exclusive 
$^{16}$O$(e,e^{\prime}pp)^{14}$C reactions, are confirmed, for both
kinematical settings here considered, also in the present
calculations.  The $^1S_0$ results shown in the figures are dominated
by the one-body current and thus by SRC, while the $\Delta$ current
gives the main contribution to the $^3P_1$ results.

One of the main results of the previous theoretical investigations is
the dominance of $^1S_0$ $pp$ knockout in the 
$^{16}$O$(e,e^{\prime}pp)^{14}$C$_{\rm g.s.}$ reaction. These
theoretical predictions have been clearly confirmed in comparison with
data. Even though the contribution of $^3P_1$ $pp$ knockout can become
important and even dominant at large values of $p_{\rm B}$, it is
clear that a TOF where only the $^3P_1$ state is included is unable to
give a reliable description of the two-proton knockout process. In
contrast, it can be seen from Fig.~\ref{cs_sup} that the cross section
calculated with the Jastrow TOF for the $^1S_0$ state is close to the
SF and also to the SM+CORR results at low values of $p_{\rm B}$, up to
$\sim 150-200$ MeV/$c$, that is just in the region where the $^1S_0$ 
contribution is dominant. For $p_{\rm B} \geq 200$ MeV/$c$ $^3P_1$ 
knockout becomes dominant with all the different treatments of the
TOF. The results with the $^3P_1$ TOF from the Jastrow TDM is however
much larger than the SF result and also larger than the SM+CORR cross
section. It can be noted that even the $^1S_0$ curve in
Fig.~\ref{cs_sup} is, at large values of the momentum, higher that the
SF result. This is an indication that SRC in the JCM produce a
stronger enhancement of the high-momentum components.

The behaviour of the pure $^1S_0$ result in the left panel of
Fig.~\ref{cs_nik} is somewhat similar to that of the SF and SM+CORR
cross sections, which appear driven by the $^1S_0$ contribution. There
are anyhow significant differences in the shape and large differences
in the size of the various results.

The cross sections calculated with the total TOF, where the $^1S_0$ and
$^3P_1$ partial components from the Jastrow TDM are combined, are
shown in the right panels of Figs.~\ref{cs_sup} and \ref{cs_nik}. It
can be seen that in both kinematical settings the $^1S_0$
component dominates at low values of $p_{\rm B}$, while the $^3P_1$
component produces a strong enhancement of the cross sections at high
momenta. The contribution of the partial $^1S_0$ component is reduced
with respect to the results in the left panels, where $^1S_0$ is fully
responsible for the knockout process. Thus, the cross sections
calculated with the total TOF from the Jastrow TDM are somewhat 
reduced at low recoil momenta. Also the contribution of the partial
$^3P_1$ component is slightly reduced with the respect to the $^3P_1$
results displayed in the left panels. The contribution of the $^3P_1$
component to the total TOF is, however, much more relevant than with
the other theoretical treatments considered in the figures and the
enhancement at high momenta turns out to be much larger. Thus, the
shape of the cross sections with the total TOF from the JCM is flatter
than with the SF and SM+CORR results. 

In comparison with the SF calculations, the cross sections with the
Jastrow TOF are lower at low recoil momenta and much larger at high
momenta, due to the larger contribution of the $^3P_1$ component in
the TOF. The SM+CORR cross sections are higher than the other results
at low recoil momenta. This is an indication of a stronger contribution
of SRC in this calculation. This contribution, however, depends on the
particular expression adopted for the correlation functions, that in
the calculations of Figs.~\ref{cs_sup} and \ref{cs_nik} is exactly the
same as in the calculation of the TDM. At high momenta the SM+CORR
cross sections remain always higher than the SF results, but generally
lower than the results given by the TOF from the JCM. 

Although obtained from a calculation of the TDM within the JCM where
only SRC are included the TOF used in our calculations are able to
reproduce the main qualitative features of the
$^{16}$O$(e,e^{\prime}pp)^{14}$C$_{\rm g.s.}$ cross sections which
were found in previous theoretical investigations and also in the
analysis of the available data. This means that the procedure suggested
in \cite{ADS+99} to calculate the TOF's from the TDM can be applied
and exploited in the study of two-nucleon knockout reactions. 

The large differences found in Figs.~\ref{cs_sup} and \ref{cs_nik}
indicate that the calculated cross sections are very sensitive to the
different approaches used and to the theoretical treatment of nuclear
structure and correlations in the TOF. It would be interesting to
apply the procedure used in this work for the calculation of the TOF's
to more refined treatments of the TDM. 

\section{Conclusions}

The results of the present work can be summarized as follows:

\begin{itemize} \item[i)] The two-nucleon overlap functions (and their
norms, the spectroscopic factors) corresponding to the knockout of two
protons from the ground state of $^{16}$O and the transition to the
ground state of  $^{14}$C are calculated using the recently
established relationship \cite{ADS+99} between the TOF's and the TDM.
In the calculations the TDM obtained within the JCM \cite{DKA+00} is
used. Though only SRC are accounted for in the Jastrow TDM, the
results can be considered as a first attempt to use an approach which
fulfils the general necessity the TOF's to be extracted from 
theoretically calculated TDM's corresponding to realistic wave
functions of the nuclear states. Of course, the quality of the results
will depend heavily on the availability of a realistic TDM
incorporating all necessary types of $NN$ correlations.

\item[ii)] The contributions of the two-proton overlap functions
corresponding to the removal of $^{1}S_{0}$ and $^{3}P_{1}$ $pp$-pairs
from $^{16}$O are calculated in two manners: 1) when each one is fully
responsible for the knockout process, and 2) when they are partial
components of the total TOF. The $^{1}S_{0}$ and $^{3}P_{1}$ results
obtained in the two manners are similar, the main difference being that
the partial components in case 2) are reduced in magnitude. The
comparison between the results of correlated (Jastrow) and
uncorrelated TOF's shows that SRC depend on both relative and CM
coordinates and affect the size and the shape of the $^{1}S_{0}$ and
$^{3}P_{1}$ overlap functions. The effects of SRC, however, are much
stronger when the two protons are in an $^{1}S_{0}$ state.

\item[iii)] The TOF's obtained from the Jastrow TDM are included in the
theoretical approach of \cite{GP91,GP97,GPA+98} to calculate the cross
section of the $^{16}$O$(e,e^{\prime}pp)^{14}$C$_{\rm g.s.}$ knockout
reaction. Calculations are performed in two different kinematics that
have been realized for the cross section measurements at NIKHEF and
MAMI. The results are compared with the cross section calculated,
within the same theoretical model for the reaction mechanism, with
different treatments of the TOF, in particular with the TOF obtained
from a calculation of the two-proton spectral function of $^{16}$O
\cite{GPA+98,GAD+96} where both SRC and LRC are included. The
calculated cross sections are very sensitive to the theoretical
treatment and different results are produced by the different TOF's.
The cross sections calculated in the present work, where the TOF's are
extracted from the Jastrow TDM, confirm the dominant contribution of
$^{1}S_{0}$ $pp$ knockout at low values of recoil momentum up to
$\simeq 150-200$ MeV/$c$. The $^{3}P_{1}$ contribution is mainly
responsible for the high-momentum part of the cross section at $p_{\rm
B} \geq 200$ MeV/$c$. 

\item[iv)] Our method is applied in the present work only to the ground
state of $^{14}$C. It can be used also for the excited states. Our
main aim was to check the practical application of all steps of the
method for a given state of the residual nucleus. Therefore, the
results obtained for the $^{16}$O$(e,e^{\prime}pp)^{14}$C$_{\rm g.s.}$
reaction, which are able to reproduce the main qualitative features of
the experimental data and of the cross sections calculated with
different treatments of the TOF's, can serve as an indication of the
reliability of the method, that can be applied in a wider range of
situations and to more refined approaches of the TDM. \end{itemize}

\section{Acknowledgments}

One of the authors (D.~N.~K.) would like to thank the Pavia Section of
the INFN for the warm hospitality and for providing the necessary
fellowship. The work was partly supported by the Bulgarian National
Science Foundation under Contracts $\Phi$-809 and $\Phi$-905.

\end{document}